\documentstyle[twocolumn,aps]{revtex}

\begin{document}

%
%

\title{Symmetry of the Gap Function in High-$T_c$ Superconductors}

\author{Mihir Arjunwadkar and D. G. Kanhere}
\address{Department of Physics, University of Poona,
         Pune 411 007, India}

\author{G. Baskaran and Rahul Basu \cite{email}}
\address{The Institute of Mathematical Sciences,
         Madras 600 113, India}

\date{December 15, 1994 (revised 30 January 1995)}

\maketitle

\begin{abstract}

One of the most debated issues related to high-$T_c$
superconductivity is the symmetry of the Cooper pair or the gap
function.
In this report, we present numerical results regarding the gap function
in strongly correlated electron systems using $t-J$ and Hubbard
models in one and two dimensions.
To this end, we use exact diagonalization to study the ground states
of 8- and 16-site clusters consisting of single or coupled layers.
We calculate a reduced two-particle density matrix in momentum space
which is a measure of the gap function.
We then analyze the eigenvectors of this density matrix, which display
the possible Cooper pair symmetries.
The eigenvector corresponding to the largest eigenvalue indicates a
vanishing gap on the Fermi surface (which is in favour of odd-gap pairing)
although $d_{x^2-y^2}$ symmetry is seen to be a very close contestant in many
of the cases.

\end{abstract}

%
%

\section{Introduction}

In the theory of superconductivity, the quantity which characterizes
the nature of pairing is the gap function.
More precisely, it is the amplitude for pairing
\begin{equation}
        \Delta_{\bf k} = \langle
                             c_{ {\bf k} \uparrow  }^\dagger
                             c_{-{\bf k} \downarrow}^\dagger
                     \rangle,
\label{gapfn}
\end{equation}
for a Cooper pair of electrons in states $({\bf k} \uparrow, -{\bf k}
\downarrow)$
in a singlet spin configuration.
This quantity is closely related to the energy gap in the single-particle
excitation spectrum of a superconductor, and has a direct bearing on
many of the experimentally measurable quantities.

As an example, consider the standard BCS $s$-wave superconductor
\cite{BCS}, which has an isotropic gap function which is
{\em even} across the Fermi surface (FS), non-zero only in a thin shell
of width $\sim \hbar \omega_D$ around $\epsilon_F$.
The energy gap $\Delta^{(e)}_{\bf k}$ is related to the gap function (1)
through the relation
\begin{equation}
        \Delta^{(e)}_{\bf k} = 2 E_{\bf k} \Delta_{\bf k},
\end{equation}
and obeys the celebrated gap equation
\begin{equation}
        \Delta^{(e)}_{\bf k} = - \sum_{{\bf k}^\prime} V_{{\bf k} {\bf
k}^\prime}
                               { \Delta^{(e)}_{{\bf k}^\prime}
                                 \over 2 E_{{\bf k}^\prime}    },
\label{gapeq}
\end{equation}
where $E_{\bf k} = \sqrt{\epsilon_{\bf k}^2 + \Delta^{(e)2}_{\bf k}}$ is the
quasiparticle energy, $\epsilon_{\bf k}$ is the single-particle
energy with respect to $\epsilon_F$, and $V_{{\bf k} {\bf k}^\prime}$
is the effective electron-electron interaction matrix element
in the reduced BCS hamiltonian.
The energy gap can be directly observed, for example, through
superconductor-normal metal or Giaver tunneling.

To understand the significance of the gap function (\ref{gapfn}),
let us examine the behaviour of the BCS wavefunction close to the FS:
\begin{equation}
   \vert BCS \rangle = \prod_{\bf k} (u_{\bf k} + v_{\bf k}
                                  c_{{\bf k} \uparrow}^\dagger
                                  c_{-{\bf k} \downarrow}^\dagger)
                                  \vert 0 \rangle
\end{equation}
which can be rewritten as
\begin{equation}
   \vert BCS \rangle = \prod_{{\bf k}}{}^\prime (u_{\bf k}^2 +
      \sqrt{2} u_{\bf k} v_{\bf k} b_{{\bf k},-{\bf k}}^\dagger +
       v_{\bf k}^2 b_{{\bf k},-{\bf k}}^\dagger b_{{\bf k},
       -{\bf k}}^\dagger) \vert 0 \rangle,
\label{BCSwf}
\end{equation}
where the product is now over only half of the ${\bf k}$-space
(e.g. $k_x > 0$ and all $k_y$ in 2D).
Here we have
\begin{equation}
   b_{{\bf k},-{\bf k}}^\dagger = \frac{1}{\sqrt{2}} (
             c_{{\bf k}\uparrow}^\dagger c_{-{\bf k}\downarrow}^\dagger
           - c_{{\bf k}\downarrow}^\dagger c_{-{\bf k}\uparrow}^\dagger )
\label{SCP}
\end{equation}
as the singlet Cooper pair creation operator at points ${\bf k}$ and
$-{\bf k}$ in ${\bf k}$-space.
It is clear that $u_{\bf k}^2$ is the {\em probability amplitude}
of finding no singlet pair with momentum $({\bf k},-{\bf k})$,
$u_{\bf k} v_{\bf k}$ is that of finding one singlet pair
(of charge $2e$), and $v_{\bf k}^2$ that of finding
two singlet pairs (of total charge $4e$).
The BCS state has identical phase relations for various configurations
of pair occupancy in ${\bf k}$-space.
That is, when the product in Eq. (\ref{BCSwf}) is expanded out,
the resulting sum has identical phase for all terms, each term
corresponding to different configurations of the $({\bf k},-{\bf k})$
occupancy.
Superconductivity can thus be thought of as a coherent charge-$2e$
fluctuating state in ${\bf k}$-space.
Since $\Delta_{\bf k}$ is non-zero only in a thin energy shell
around the FS, the coherent $2e$ charge fluctuation is concentrated
around the FS.
(Note that this coherence in ${\bf k}$-space
results in phase coherence among the Cooper pairs in real space also).
Away from the shell, we either have a completely filled band (inside the
FS) or a completely empty band (outside the FS) and
hence no charge fluctuations.
The gap function (\ref{gapfn}), for the BCS wavefunction (\ref{BCSwf}),
reduces to
\begin{equation}
        \Delta_{\bf k} = u_{\bf k} v_{\bf k},
\end{equation}
and is thus a measure of coherent charge fluctuations
in ${\bf k}$-space.

In the context of the high-$T_c$ cuprates, it is generally believed that
the gap function is not of simple BCS type ($s$-wave and even across ${\bf
k}_F$).
Some of the recent experiments have suggested that the
gap is highly anisotropic, possibly with the $d_{x^2-y^2}$ symmetry.
\cite{dwave_ARPES}
Yet another proposal is that of odd-gap pairing,
which is a favoured mode of pairing in the presence of a
strong, purely repulsive interaction. \cite{Mila}
A recent proposal in the context of the interlayer tunneling mechanism
is that of anisotropic $s$-wave pairing, which has deep minima but
no actual nodes on the FS. \cite{anisotropic_s}
However, in spite of intense activity, both on theoretical and
experimental sides, the issue of the symmetry of the gap function
still remains unresolved.
The importance of this issue cannot be overestimated, since the
gap function is a quantity which, on the one hand, has direct experimental
consequences, and on the other, may offer a unique signature
pointing to the right kind of mechanism for high-$T_c$ superconductivity.

The aim of the present work is to investigate the behaviour of
$\Delta_{\bf k}$ using exact diagonalization, especially with
reference to its nodal structure.
We would like to find out the most likely pairing symmetries,
both angular and radial, for clusters consisting of single or
coupled chains and planes.
In particular, we are interested in the $d_{x^2-y^2}$ symmetry
and odd-gap pairing.
In what follows, we first present a brief summary of the various
proposed gap functions, especially $d_{x^2-y^2}$ and odd-gap
pairing (section II).
We then present our numerical procedure in section III and the
results in section IV.
Finally, conclusions are presented in section V.

%
%

\section{$d_{x^2-y^2}$, Odd-Gap and Other Pairing Symmetries}

In order to see, on general grounds, why gap functions with zeros
are preferable as candidates for pairing in high-$T_c$ superconductors,
let us consider the one-band large-$U$ Hubbard model or the
$t-J$ model.
These models are believed to contain the
essential low-energy physics of high-$T_c$ superconductivity.
Let us begin by noting that in the large-$U$ Hubbard model, the on-site
pairing amplitude
$\Delta_{ii} = \langle
                       c_{i \uparrow}^\dagger c_{i \downarrow}^\dagger
               \rangle$
is diminished considerably on account of the restriction on double
occupancy.
In the $t-J$ model, of course, double occupancy is completely
projected out.
Thus $\Delta_{ii} = 0$, which reduces, in ${\bf k}$-space, to
\begin{equation}
  \Delta_{ii} = \sum_{\bf k} \langle
                          c_{ {\bf k} \uparrow  }^\dagger
                          c_{-{\bf k} \downarrow}^\dagger
                       \rangle
              \equiv \sum_{\bf k} \Delta_{\bf k} = 0,
\label{gapsum}
\end{equation}
where we have used the pairing condition
$\langle c_{{\bf k} \uparrow}^\dagger
         c_{{\bf k}^\prime \downarrow}^\dagger
 \rangle
 = \Delta_{\bf k} \delta_{{\bf k},-{\bf k}^\prime}$.
This is a constraint on the gap function, arising from
a strong real-space repulsion, and implies the presence
of zeros in the gap function.

The constraint, however, is a global constraint, and can
be satisfied in several ways.
One possibility is that of odd-gap pairing, in which case
the gap function vanishes on the {\em entire} Fermi surface,
and changes sign across the FS.
Yet another possibility is that of
a simple $d_{x^2-y^2}$ superconductor on a 2D square lattice, for which
the gap function vanishes along the lines $k_x = \pm k_y$
which intersect the FS at exactly four points.
We give below a brief account of these two and some of the other
proposed pairing states.

\subsection{Odd-Gap Pairing}

The possibility of odd-gap pairing, in the presence of
a dominantly repulsive interaction,
was anticipated by Cohen \cite{Cohen} many years ago.
The case for odd-gap pairing, in the context of high-$T_c$
superconductivity, was put forward by Mila and Abrahams
\cite{Mila}, within the framework of the weak coupling
BCS theory.
They have shown that an {\em even} solution to the BCS gap
equation (\ref{gapeq}) exists only so long as the effective
electron-electron interaction is dominantly attractive,
albeit weak.
This is the case with the standard BCS model of conventional
superconductivity, where the weak attraction arises due to
electron-phonon coupling.
In the presence of a repulsive interaction, however,
such a solution ceases to exist when the repulsive part
of the interaction starts dominating,
and an unusual solution is shown to exist which is {\em odd}
across the FS.
For example, in the toy models considered in \cite{Cohen}
as well as \cite{Mila}, the energy gap is proportional to
$(\epsilon - \epsilon_F)$, which is clearly odd across the FS.
Moreover, an odd gap is independent of the {\em magnitude} of
the repulsion, provided that it is the repulsion which dominates.
This condition is met in the large-$U$ Hubbard model,
and odd-gap pairing is thus a candidate for pairing in the
high-$T_c$ superconductors.
Mila and Abrahams have further shown that
certain key features of the tunneling density of states in
the cuprates, such as the temperature-insensitive peak, can
be naturally explained on the basis of odd-gap pairing.
This feature is in contrast with $s$-wave as well as $d_{x^2-y^2}$
pairing, in which the peak shifts towards $\epsilon_F$ as
$T \rightarrow T_c$.

Anderson \cite{PWA} has also pointed out in the context
of the RVB theory that the gap function vanishes on, and
changes sign across, the FS.
For example, in the RVB mean field theory at half filling,
\cite{BZA} the gap function is constant and changes sign across
the pseudo FS defined by $\cos k_x + \cos k_y = 0$ and satisfies
the constraint (\ref{gapsum}).
For the doped case, however, the analysis is difficult and
Anderson did not provide any argument as to why the gap
should vanish on the FS away from half-filling. \cite{BZA}
In the latter case, it is argued that strong correlations in real space
lead to a suppression of coherent charge fluctuations close
to the FS, \cite{speculate} in the sense of Eq. (\ref{BCSwf}),
as follows.
Strongly correlated electrons in 1D, described by a
large-$U$ Hubbard model, have certain unique features close to the FS.
It is well known that in the 1D Hubbard model there is singular
forward scattering between two electrons with opposite spins close
to the FS.
This leads to a finite phase shift \cite{SFS} at the FS and the
consequent failure of the Fermi liquid theory, resulting in the
vanishing of the discontinuity in $n_{\bf k}$ at the FS
(Luttinger liquid behaviour).
It also implies an effective hard-core repulsive pseudopotential
between electrons with opposite spins close to the FS.
Thus no two electrons close to the FS, with opposite spins,
can have the same momentum, thereby making ${\bf k}$-points close
to the FS essentially singly occupied.
Coherent pair fluctuations are thus unlikely to develop on or very
close to the FS, but are not forbidden away from the FS.
It is therefore likely that the line of zeros of $\Delta_{\bf k}$
implied by the global constraint (Eq. (\ref{gapsum})) will coincide
with the FS.

Whether these arguments carry over to 2D or not is still unclear.
It is believed that if the constraint of no double occupancy in
real space leads to singular forward scattering and the consequent
failure of Fermi liquid theory in 2D (in the spirit of
Anderson's tomographic Luttinger liquid picture \cite{TLL}),
a gap function which vanishes on the entire FS is a natural
candidate for pairing.

Odd-gap pairing has so far found very feeble experimental
support.
However, a very recent work \cite{odd_ARPES} shows ARPES data
in Bi2212 cuprates in which the gap {\em does not vanish} along
$k_x = k_y$, but vanishes on either side of this line.
This can be interpreted as a possible indication that the
nodal lines in the gap coincide with the FS.

\subsection{$d_{x^2-y^2}$ Pairing}

The $d_{x^2-y^2}$ paired state is characterised by a gap function
of the form
\begin{equation}
        \Delta_{\bf k} = \Delta_d ( \cos k_x - \cos k_y ).
\end{equation}
Support to $d_{x^2-y^2}$ pairing comes from numerous sources.
Theoretical analysis of motion of holes in an antiferromagnetic
background shows that exchange of spin fluctuations can induce
singlet pairing of $d_{x^2-y^2}$ symmetry, for which
hole-hole interaction is attractive. \cite{dwave_T}
Although antiferromagnetic fluctuations are rather subdued
in the doped case in comparison with half-filling, it
is reasonable to expect $d_{x^2-y^2}$ pairing as a possibility.

On the experimental side, numerous experiments have been
performed recently, which seem to support $d_{x^2-y^2}$ pairing.
London penetration depth data from Hardy {\em et al.}
\cite{dwave_LPD} shows a low-temperature linear temperature
dependence which is in agreement with a $d_{x^2-y^2}$ superconductor,
and very much in contrast with the BCS $s$-wave case
(which shows an exponential behaviour).
An experimental result strongly supporting $d_{x^2-y^2}$
pairing is the $T^3$ variation of Cu NMR relaxation rate
\cite{dwave_NMR}.
ARPES measurements on Bi2212 compunds
by Shen {\em et al.} \cite{dwave_ARPES} indicate lines
of zeros compatible with a $d_{x^2-y^2}$ state, and
incompatible with an extended-$s$ state, although these
measurements are unable to rule out pairings such as $s+id$
or an anisotropic $s$-wave (next section).
Certain other experiments, which support $d_{x^2-y^2}$ pairing, but
cannot rule out the possibility of other kinds of pairing
include SQUID experiments on YBCO \cite{dwave_SQUID} and electronic
Raman scattering measurements on Bi2212 compounds. \cite{dwave_Raman}

On the numerical side, studies in favour $d_{x^2-y^2}$ pairing include
an exact diagonalization study of possible pairing symmetries
by Riera and Young \cite{dwave_Riera} (on which we comment in the
next section).
Dagotto and Riera \cite{dwave_Dagotto} find bound hole pairs with
$d_{x^2-y^2}$ symmetry near half-filling and on the verge of phase
separation, although the value of $J/t \sim 3$ is rather too large
in comparison with the experimentally relevant parameter range
$J/t \sim 0.3$.
Another notable work \cite{dwave_Maekawa} calculates an anomalous
Green's function for one and two holes using Lanczos technique,
and finds signals of $d_{x^2-y^2}$ pairing over a wide range of parameter
$J/t$ (not specified).

\subsection{Other Proposed Pairing Symmetries}

A variational Monte Carlo study by Li {\em et al.}
\cite{dwave_vmc} supports $s+id$ symmetry rather than
a pure $d_{x^2-y^2}$.
This calculation compares RVB variational states with
extended-$s$, $d_{x^2-y^2}$, and mixed states $s+d$ and $s+id$
symmetries.
The authors show that the mixed states are energetically
preferred over pure $s$ or $d_{x^2-y^2}$ states, and that the
$s+id$ state overcomes finite-size effects at
a lattice size smaller than that required by the $d_{x^2-y^2}$ state.
The authors also argue that the Knight shift data in the
YBCO cuprates is better explained by $s+id$ rather than
pure $d_{x^2-y^2}$ pairing.
Wheatley \cite{dwave_JW} has also analyzed the stability
of various mixed states involving $d_{x^2-y^2}$.

Chakravarty {\em et al.} \cite{anisotropic_s} have recently
proposed an anisotropic $s$-wave pairing state
based on the interlayer tunneling mechanism \cite{WHA}.
This state has no actual nodes on the FS although deep minima occur,
and the gap is highly anisotropic.
Since ARPES results, such as \cite{dwave_ARPES}, are unable to
detect the sign of the gap function (as well as the presence of a
node unambiguously owing to the current energy resolution),
they are consistent with such an anisotropic $s$-wave pairing as well.
Interlayer tunneling is also responsible for a temperature-dependent
anisotropy in the gap, as shown by Muthukumar and Sardar \cite{Muthu},
which is incompatible with the $d_{x^2-y^2}$ scenario.
Such a temperature-dependent anisotropy has been observed in the
ARPES data by Ma {et al.} \cite{Ma_ARPES}

%
%

\section{Numerical Procedure}

In our exact diagonalization computation, since we have chosen to work
in a number-conserving basis, it is not possible to evaluate
$\Delta_{\bf k}$ directly. \cite{directly}
We thus define a correlation function
$A_{{\bf k},{\bf k}^\prime}$ as
\begin{equation}
   A_{{\bf k},{\bf k}^\prime} = \langle
                                b_{{\bf k},-{\bf k}}^\dagger
                                b_{{\bf k}^\prime,-{\bf k}^\prime}
                        \rangle,
\label{dm}
\end{equation}
where the average $\langle \ldots \rangle$ is the
expectation value in the ground-state of a finite cluster,
obtained via exact diagonalization.
This quantity is clearly a measure of coherent pair fluctuations
between states $({\bf k},-{\bf k})$ and
$({\bf k}^\prime,-{\bf k}^\prime)$ in the ground state.

We find the gap function $\Delta_{\bf k}$ by diagonalizing this
two particle reduced density matrix (\ref{dm}), which has
the eigenfunction decomposition
\begin{equation}
  A_{{\bf k},{\bf k}^\prime} = \sum_\alpha \lambda_\alpha
                                   \Delta_\alpha({\bf k})
                                   \Delta^\ast_\alpha({\bf k}^\prime).
\end{equation}
Here, $\lambda_\alpha$ and $\Delta_\alpha({\bf k})$ are the
$\alpha$ th eigenvalue and eigenfunction of the
$N \times N$ matrix $A_{{\bf k} {\bf k}^\prime}$.
Here $N$ is the number of points in the Brillouin zone and
the index $\alpha$ orders $\lambda_{\alpha}$'s as
$\lambda_1 \geq \lambda_2 \geq \lambda_3 \geq \ldots$.
Superconducting ODLRO is signalled \cite{Yang} by a macroscopic
separation of the largest eigen value $\lambda_1$ from the next
one $\lambda_2$, i.e. $\lambda_1 - \lambda_2 \approx N$.
The required gap function $\Delta_{\bf k}$ is the eigenvector
$\Delta_1({\bf k})$ corresponding the largest eigen value $\lambda_1$,
which is the Cooper pair state with least energy for which
condensation will take place, whereas the other eigenvectors
represent the various excited states of the Cooper pair.
For example, for the standard BCS ground state,
$\lambda_1 = N$, $\Delta_1({\bf k}) = u_{\bf k} v_{\bf k}$ and
$\lambda_\alpha = 0$, for $\alpha = 2, 3 \ldots N$.

A similar procedure was adopted by Riera and Young \cite{dwave_Riera}
in the context of the $t-J$ model with the three-site term.
However, the authors argue that from the expression (\ref{dm})
translated into real space,
\begin{equation}
   A_{{\bf k},{\bf k}^\prime} = {1 \over N^2} \sum_{klmn}
                                e^{  i {\bf k}        \cdot ({\bf R}_k - {\bf
R}_l)}
                                e^{- i {\bf k}^\prime \cdot ({\bf R}_m - {\bf
R}_n)}
                                \langle b_{kl}^\dagger b_{mn} \rangle,
\label{dmrs}
\end{equation}
(where $b_{mn}^\dagger = {1 \over \sqrt{2}} (
             c_{m \uparrow}^\dagger   c_{n \downarrow}^\dagger
           - c_{m \downarrow}^\dagger c_{n \uparrow}^\dagger   )$
creates a spin-singlet pair of electrons at sites $m, n$),
one should retain only those terms which have all the four indices
$k, l, m, n$ distinct, since it is only these terms which correspond
to true pairing terms, rather than mere charge and spin correlations.
The authors {\em post facto} justify this procedure
on the basis that the $d_{x^2-y^2}$ symmetry starts dominating
very close to where the two-hole binding energy becomes negative,
as a function of $J$.
We do not agree with this argument, since there is no clear cut way
to justify such a removal {\em a priori}, in our calculation,
since we are calculating a quantity in ${\bf k}$-space;
Eq. (\ref{dmrs}) then dictates that all terms be retained in the
Fourier transform.
Indeed, such removal in the $U < 0$ case leads to an incorrect
symmetry, whereas the gap function is known to be of a simple
BCS type, {\em i. e.} an even function across the FS
(with an $s$-wave symmetry in 2D).
This is illustrated in Fig. 1 in which we plot the gap function
corresponding to the highest eigenvalue for an 8-site Hubbard
chain with $U = -5$, with and without following this
removal of terms.
Without the removal of terms, we see the expected behaviour,
whereas with the removal, we observe a sign change across ${\bf k}_F$.
On a $\sqrt{8} \times \sqrt{8}$ Hubbard plane at $U = -5$, with the removal of
terms,
we have observed a double sign change across the FS, indicating
two nodal lines encircling the FS on either side,
whereas without such removal, we see the expected BCS-like symmetry.

We shall thus present results {\em without} the removal
of these ``on-site'' terms.
Note that with removal, we indeed obtain similar results as
Riera and Young. \cite{dwave_Riera}
We would also like to point out that unlike their work,
the focus of our work is the nodes in the gap function.
We shall refer to the gap function corresponding to the highest
eigenvalue of the density matrix (\ref{dm}) as the ``topmost''
gap function,
and the term eigenspectrum will be used to refer to the eigenvalue
spectrum of the density matrix.

%
%

\section{Results and Discussion}

The clusters used for these calculations are single-layer
clusters of 8 (Hubbard) and 16 ($t-J$) sites, and clusters
with two 8-site $t-J$ layers coupled by an interlayer
term of the form
\begin{equation}
      H_\perp = -t_\perp \sum_{ilm\sigma}
                (1 - n^l_{i-\sigma})
                   c^{l\dagger}_{i\sigma}
                   c^m_{i\sigma}
                (1 - n^m_{i-\sigma})
\label{ilc}
\end{equation}
where $l,m$ are layer indices (= 1,2), $i$ is the site
index within a layer and $\sigma$ is the spin label.

\subsection{Results in One Dimension}

We first present the gap function for the 1D clusters.
Fig. 2 displays the topmost gap function for $U = 10$ at
various fillings.
A clear change of sign is visible across the corresponding
Fermi point (${\pi \over 2}$ at half filling, ${\pi \over 4}$
for 2 and 4 holes), indicating the presence of a node
near ${\bf k}_F$.
Similar behaviour is observed for all $U > 0$, and
over a range of fillings (0, 2, 4 and 6 holes).

The results for a 16-site $t-J$ chain are presented in Fig. 3 and 4,
and are representative of the parameter range $J = 0.08 - 0.32$.
Fig. 3 presents eigenspectrum of the density matrix
for various cases: half-filling at $J = 0.24$ (1),
2 holes at $J = 0.32, 0.24, 0.16, 0.08$ (2)-(5).
All these spectra display a prominant, well-separated
largest eigenvalue, indicating the possibility of
ODLRO (as far as a finite size calculation can reveal).
For comparison, we have also plotted the eigenspectrum
for a antiparallel triplet Cooper pair density matrix
in Fig. 3 (6) for $J=0.24$, 2 holes, which does not
show a well-separated eigenvalue.
Fig. 4 displays the topmost gap function at half-filling
and 2 holes, where a clear sign change across ${\bf k}_F$
indicates a node near ${\bf k}_F$.

Finally, Fig. 5 plots the topmost gap function for a
{\em single} 8-site chain in a system of two 8-site $t-J$
chains coupled by an interlayer coupling term (\ref{ilc}),
for 2 holes with $J = 0.24$ and $t_\perp = 0.05-0.7$.
We see again a clear node near ${\bf k}_F={\pi \over 2}$
very similar to the case of a single 16-site $t-J$ chain.

\subsection{Results in Two Dimensions}

We now turn to the more interesting case of 2D clusters.
These results are obtained for $\sqrt{8} \times \sqrt{8}$ Hubbard, $4 \times 4$
$t-J$ and
coupled $\sqrt{8} \times \sqrt{8}$ $t-J$ planes.
The results are representative of a range of
parameter values and fillings as specified below.
$\sqrt{8} \times \sqrt{8}$ Hubbard cluster: all $U > 0$ (0, 2 and 4 holes);
$4 \times 4$ $t-J$ plane: $J = 0.08, 0.16, 0.24, 0.32$ (0 and 2 holes);
coupled $\sqrt{8} \times \sqrt{8}$ $t-J$ planes: $J = 0.24,0.32$, $t_\perp =
0.05-0.7$
(0 and 2 holes).
To bring about the symmetry features of a gap function,
we shall display the numerical values of the coefficients
of the corresponding eigenvector (of the density matrix (\ref{dm}))
at all ${\bf k}$-points in the Brillouin zone.
The FS (and the nodal lines of $d_{x^2-y^2}$, wherever required)
will be indicated by dashed lines.

We first discuss the characteristics of the eigenspectra.
In all the cases presented, the eigenspectrum displays
a separated largest eigenvalue, although not as well-separated
as in the 1D case.
Fig. 6 (1)-(3) are eigenspectra for 2 holes, respectively
for a $\sqrt{8} \times \sqrt{8}$ Hubbard cluster with $U = 10$, $4 \times 4$
$t-J$ plane
with $J = 0.24$ and coupled $\sqrt{8} \times \sqrt{8}$ $t-J$ planes with $J =
0.24,
t_\perp = 0.2$.
The most striking feature of these results is the close
interplay between the odd-pairing and the $d_{x^2-y^2}$ gap
functions.
The odd-paired state is the topmost for all the three
clusters, with the next dominant state as the $d_{x^2-y^2}$
state on $4 \times 4$ and coupled $\sqrt{8} \times \sqrt{8}$ clusters.
For the $\sqrt{8} \times \sqrt{8}$ Hubbard cluster with 2 holes, however,
the $d_{x^2-y^2}$ state is quite low down in the eigenspectrum.
These two states, for a $4 \times 4$ plane, are displayed in Fig. 7.

\subsection{Discussion}

The overall picture that emerges out of this study is as
follows.
In 1D and 2D, the gap function shows the presence of a node near
or at ${\bf k}_F$, indicating an odd-paired behaviour.
The $d_{x^2-y^2}$ state is often seen as the next dominant state.
Quite clearly, the numerical results seem to indicate the
presence of a nodal ``surface'' (GNS) in the gap function
closely following the FS.
Recall that in the RVB mean-field theory at half-filling,
the GNS coincides with the FS in 2D (also seen in our numerical
results (not shown)).
It may be argued that as doping increases from zero, the FS shrinks,
dragging part of the GNS with it.
\cite{speculate}
The non-overlap of the GNS with the FS in some parts in the BZ
is possibly caused due to the enhancement of the interlayer pair
tunneling matrix element $\frac{t^2_{\perp}(k)}{t}$ in those
directions in $\bf k$-space.
For example, as emphasized by Chakravarty {\em et al.} \cite{anisotropic_s},
$\frac{t^2_{\perp}(k)}{t}$ is largest in the $(0,\pi)$ and $(\pi,0)$
directions which enhances pairing in those regions of the FS by keeping
the GNS away.
Fig. 8 incorporates this idea as well as the eight nodal
points of the recent ARPES data \cite{odd_ARPES}.
Based on the present numerical work, we believe that it may
be possible to construct novel scenarios where such a nodal
surface is also consistent with $d_{x^2-y^2}$ pairing.
This and such other aspects are discussed elsewhere.
\cite{speculate}

The major handicap of any exact diagonalization calculation
is the cluster size restriction.
It is not possible for us to rule out finite size effects,
but the features we have observed are consistent within the
sizes and geometries considered.
It is indeed possible that finite size effects are more
serious when symmetry issues are concerned,
in comparison with, for example, the finite size effects
in simple spin, charge or pairing correlation functions.
\cite{dwave_vmc}

Note that we do not interpret the present results as indicative
of the presence of superconducting ODLRO in purely 2D systems.
However, it does indicate that these pairing correlations show an
odd-paired behaviour across the FS, a feature which seems to persist
even in the presence of an interlayer coupling term.

%
%

\section{Summary}

To summarise, we have investigated the behaviour of the
gap function $\Delta_{\bf k}$ across the Fermi surface
in the doped $t-J$ and large-$U$ Hubbard clusters in one
and two dimensions, with special reference to odd-gap
and $d_{x^2-y^2}$ pairing.
For this purpose, we diagonalize a reduced two-particle
(Cooper pair) density matrix computed through exact
diagonalization.
The largest eigenvalue of this density matrix is well-separated
from the rest, indicating the possibility of ODLRO.
The corresponding eigenvector, the Cooper pair wavefunction
with least energy, reflects the symmetry of the gap function
$\Delta_{\bf k}$.
The results on 8- and 16-site single or coupled
chains and planes clearly indicate a change of sign of
$\Delta_{\bf k}$ across ${\bf k}_F$, indicating the presence of
a node close to the entire FS, a signature of odd-pairing
behaviour.
This state, on single and coupled planes, in addtion,
is isotropic, indicating an angular symmetry which is
of the $s$-wave type.
We also observe a close interplay between the odd-paired
and $d_{x^2-y^2}$ states in the 2D systems.

%
%

\acknowledgements

We would like to thank V. N. Muthukumar for many useful
discussions.
M. A. thankfully acknowledges R. E. Amritkar for his
critical comments,
the Council for Scientific and Industrial Research, New Delhi for
financial support, and the Center for Development of Advanced
Computing (C-DAC), Pune, for computing facility.
Partial financial assitance was provided under Project No.
SP/S2/M-47/89 by the Department of Science and Technology and DST
project SBR 32 of the National Superconductivity Programme.

%
%

%
%

{\bf Figure Captions}

\begin{enumerate}

\item
Topmost gap function for an 8-site Hubbard chain with 2 holes
at $U = -5$,
{\em with} and {\em without} the removal of terms from the gap
expression (\ref{dmrs}); ${\bf k}_F = {\pi \over 4}$.
Notice the sign change across the FS {\em with} the removal of terms.

\item
Topmost gap functions for an 8-site Hubbard chain with $U = 10$,
for 0, 2 and 4 holes; ${\bf k}_F = {\pi \over 2}, {\pi \over 4},
{\pi \over 4}$ respectively.

\item
Eigenvalue spectra of the density matrix for the 16-site $t-J$ chain.
(1) Half-filled, $J=0.24$,
(2)-(5) 2 holes, $J=0.32,0.24,0.16,0.08$,
(6) {\em triplet antiparallel} eigenspectrum for 2 holes, $J=0.24$.

\item
Topmost gap functions for a 16-site $t-J$ chain at half filling
($J = 0.24$) and 2 holes ($J = 0.08 - 0.32$).
${\bf k}_F = {\pi \over 2}, {3\pi \over 8}$ for 0 and 2 holes
respectively.

\item
Topmost gap function for a single 8-site $t-J$ chain
in a coupled system of two chains with 2 holes,
for values of $t_\perp = 0.05-0.7$, $J = 0.24$.
${\bf k}_F = {\pi \over 2}$ for $k_y=0$.

\item
Eigenvalue spectra of the density matrix for the three
2D clusters at half filling and for 2 holes.
(1) $\sqrt{8} \times \sqrt{8}$ Hubbard plane, 2 holes, $U = 10$;
(2) $4 \times 4$ $t-J$ plane, 2 holes, $J = 0.24$;
(3) coupled $\sqrt{8} \times \sqrt{8}$ $t-J$ planes, 2 holes, $J = 0.24,
t_\perp = 0.05-0.7$.
The labels ``o'' and ``d'' near a level respectively stand for
odd-paired and $d_{x^2-y^2}$ states.
A numeral near a level indicates the degeneracy.

\item
The topmost two degenerate gap functions for a $4 \times 4$
plane with 2 holes, $J=0.24$.
Points marked by squares lie within one Brilliouin zone,
diamonds belong to the neighbouring ones.
The FS and the nodes of the $d_{x^2-y^2}$ state are
indicated by dashed lines.
The numbers above and below a square are
the coefficients of the gap function respectively for
the topmost (odd-paired) and next ($d_{x^2-y^2}$) states.

\item
A suggested form of the GNS with s-symmetry.
The solid line is the FS and the dotted line is the GNS.
The + and - symbols indicate relative
signs of the gap function across the GNS.

\end{enumerate}


\begin{references}
\bibitem{scaling}
More precisely, the largest eigenvalue of a density matrix
with ODLRO scales as the number of particles in the system.
However, for power-law correlations, the scaling will be
with a power smaller than one.
\end{references}
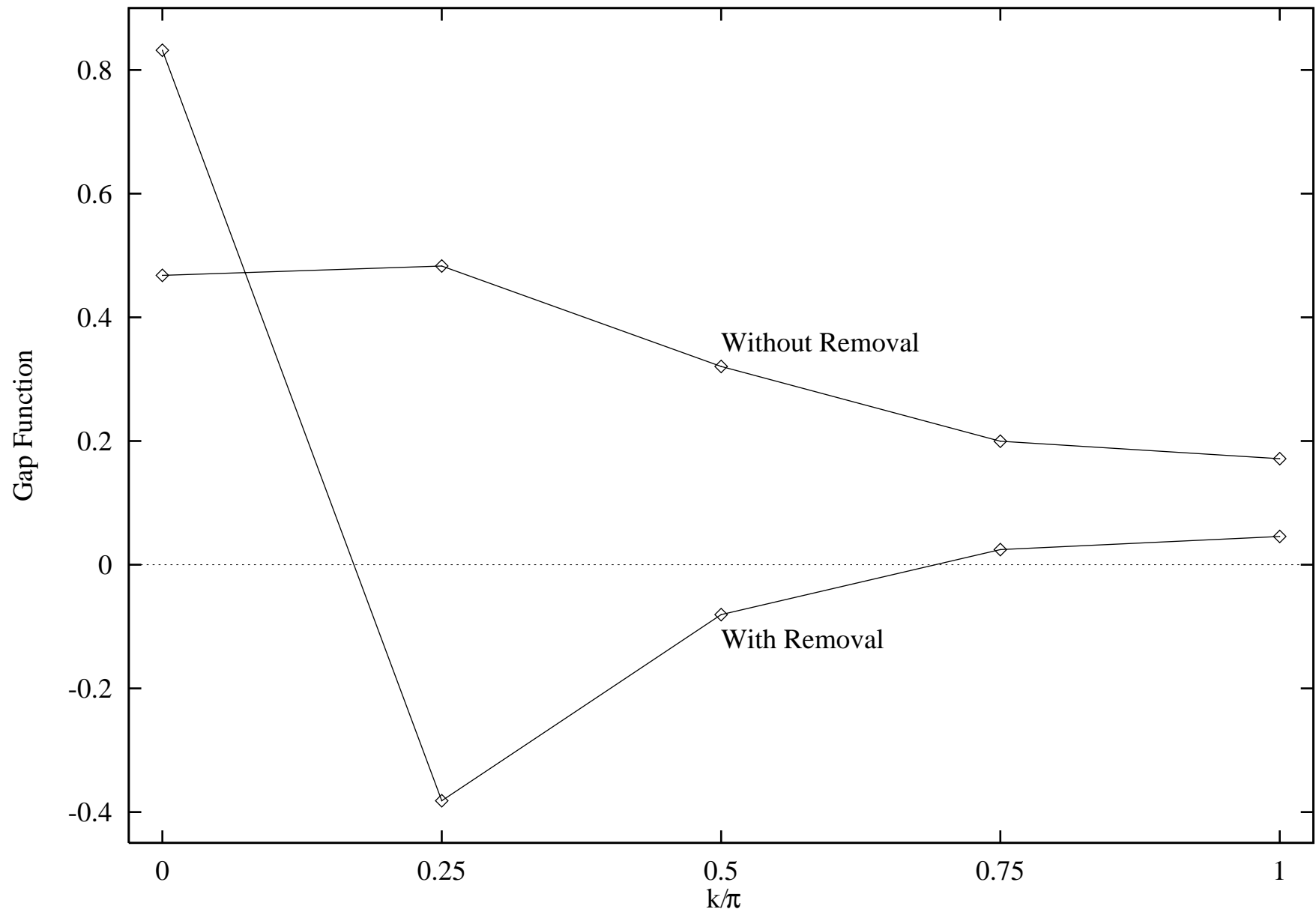
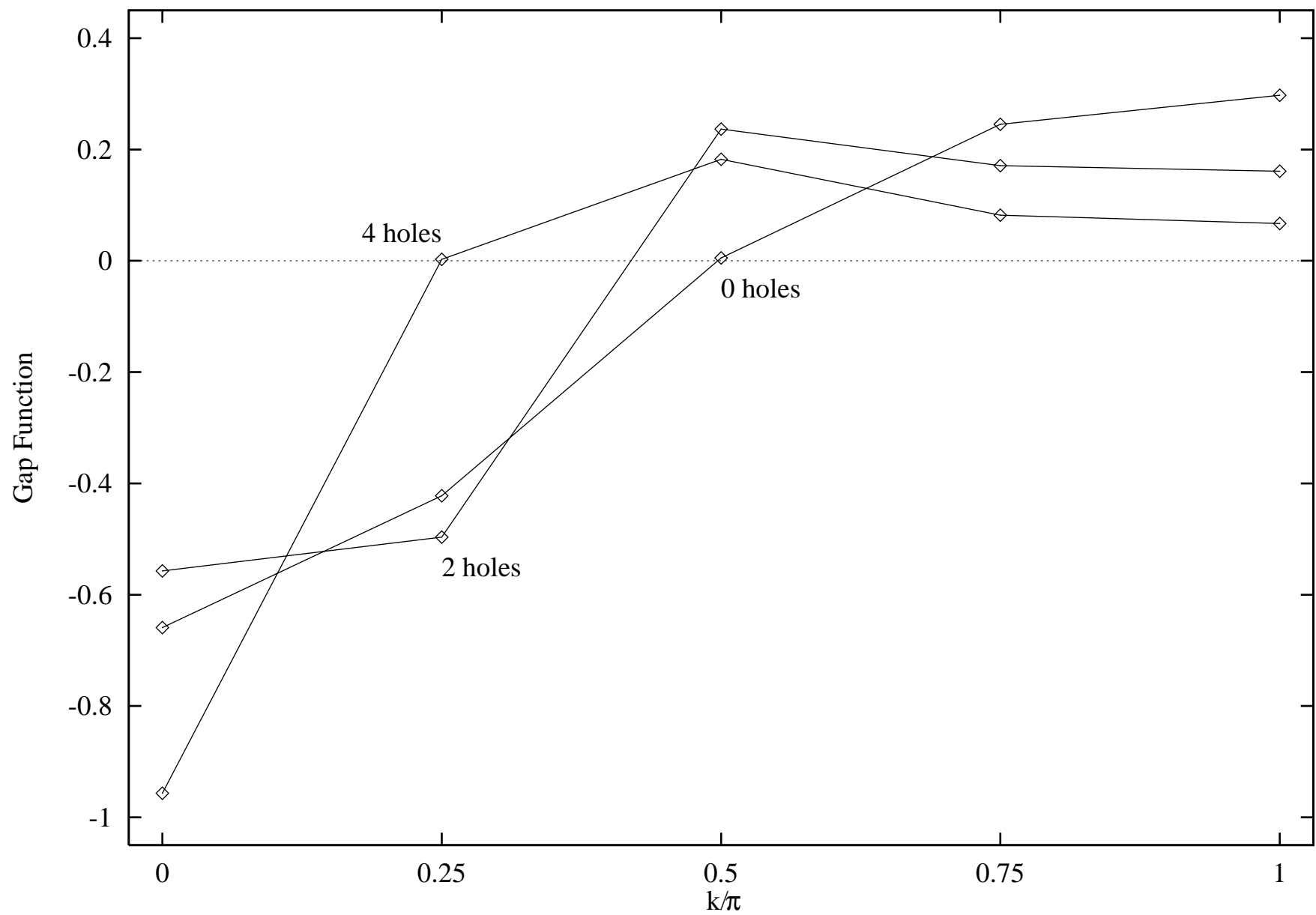
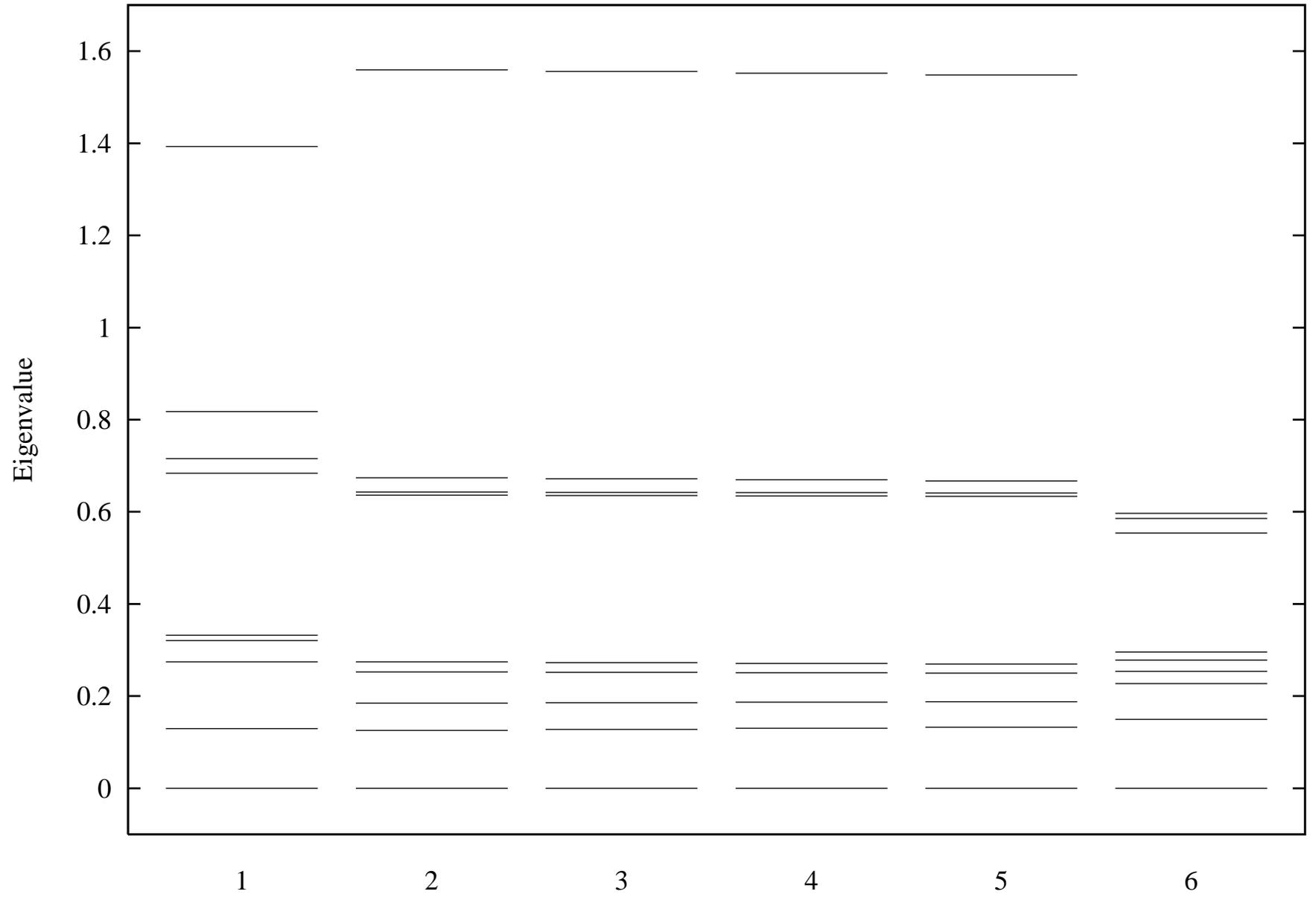
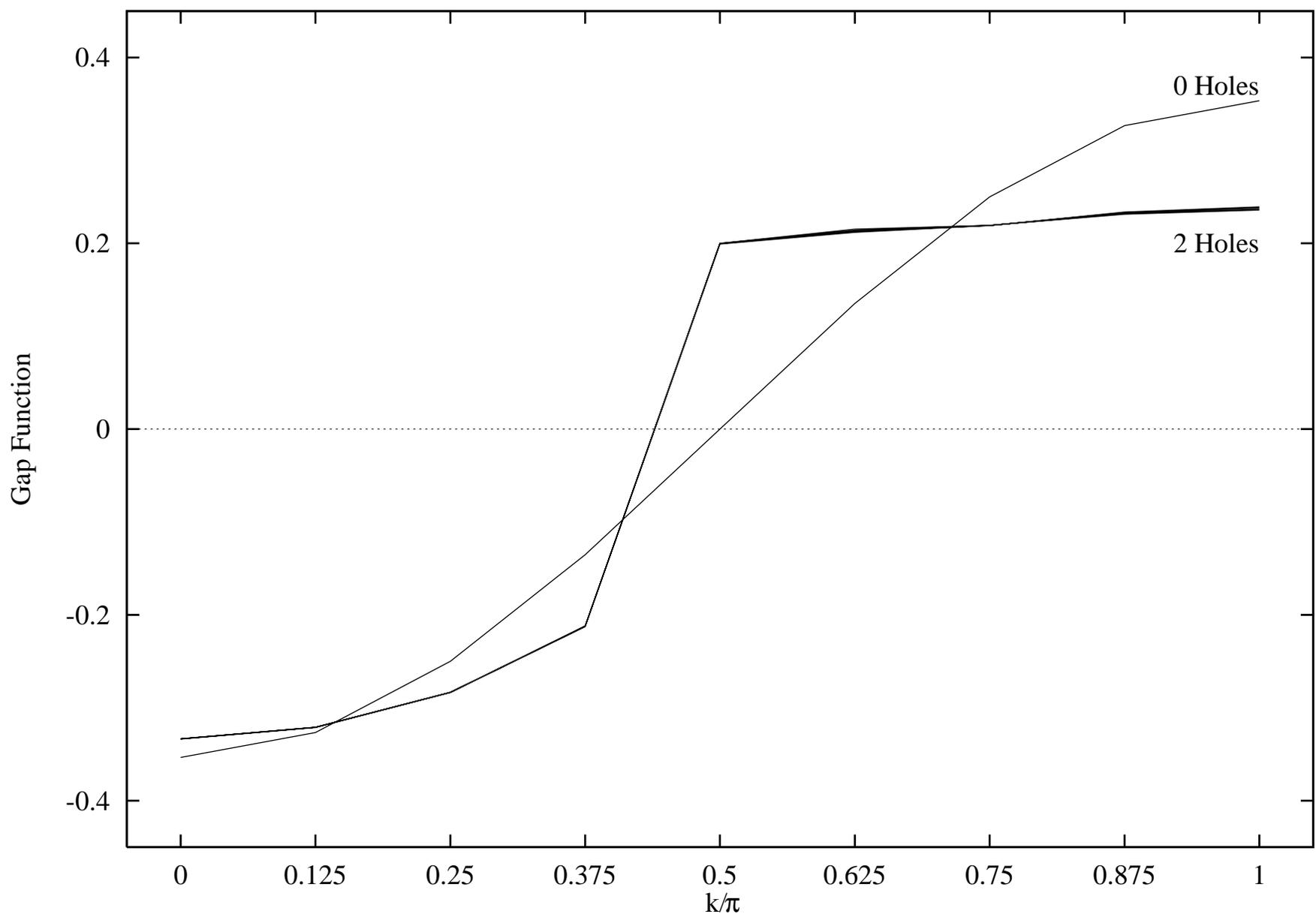
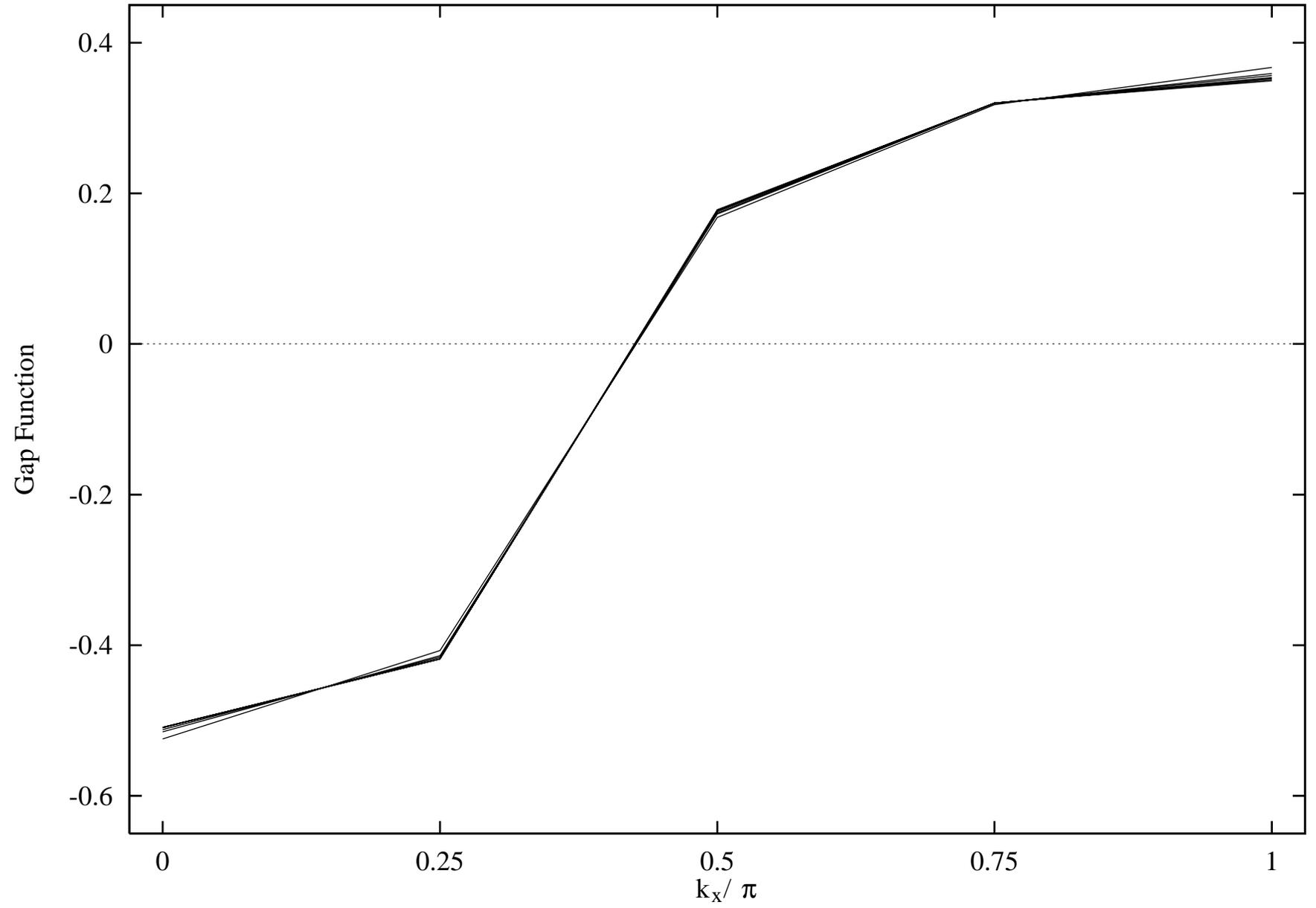
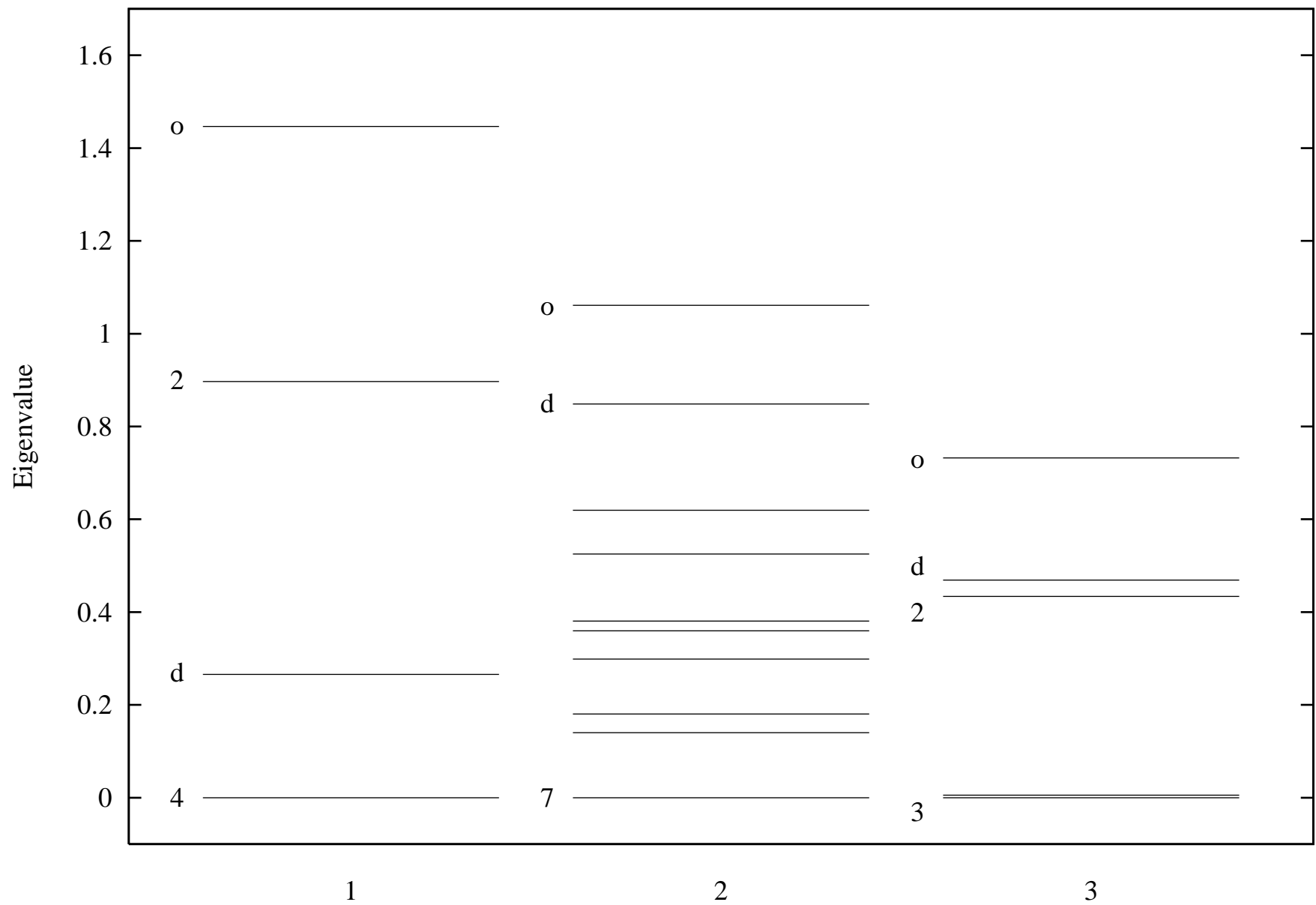
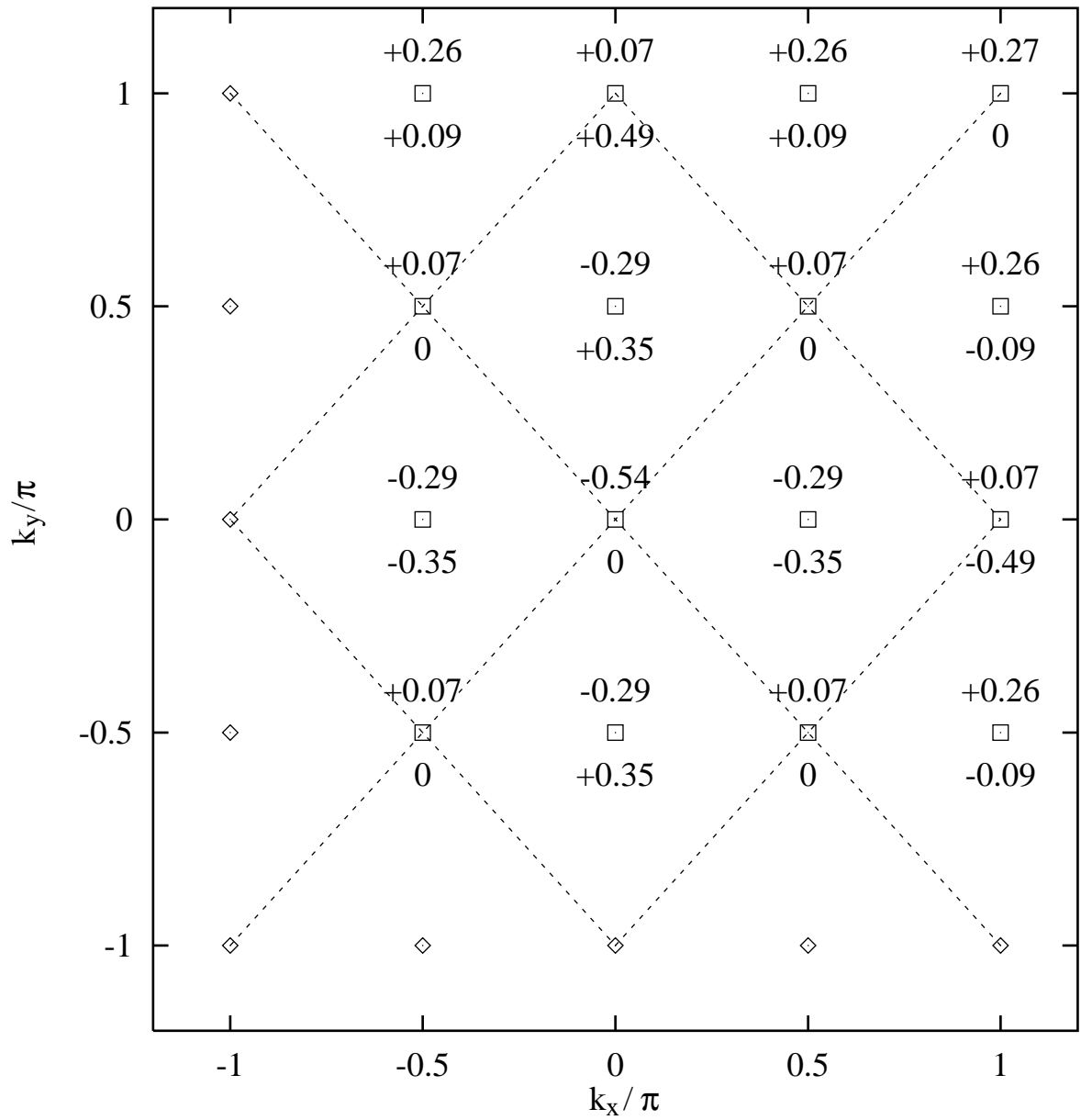
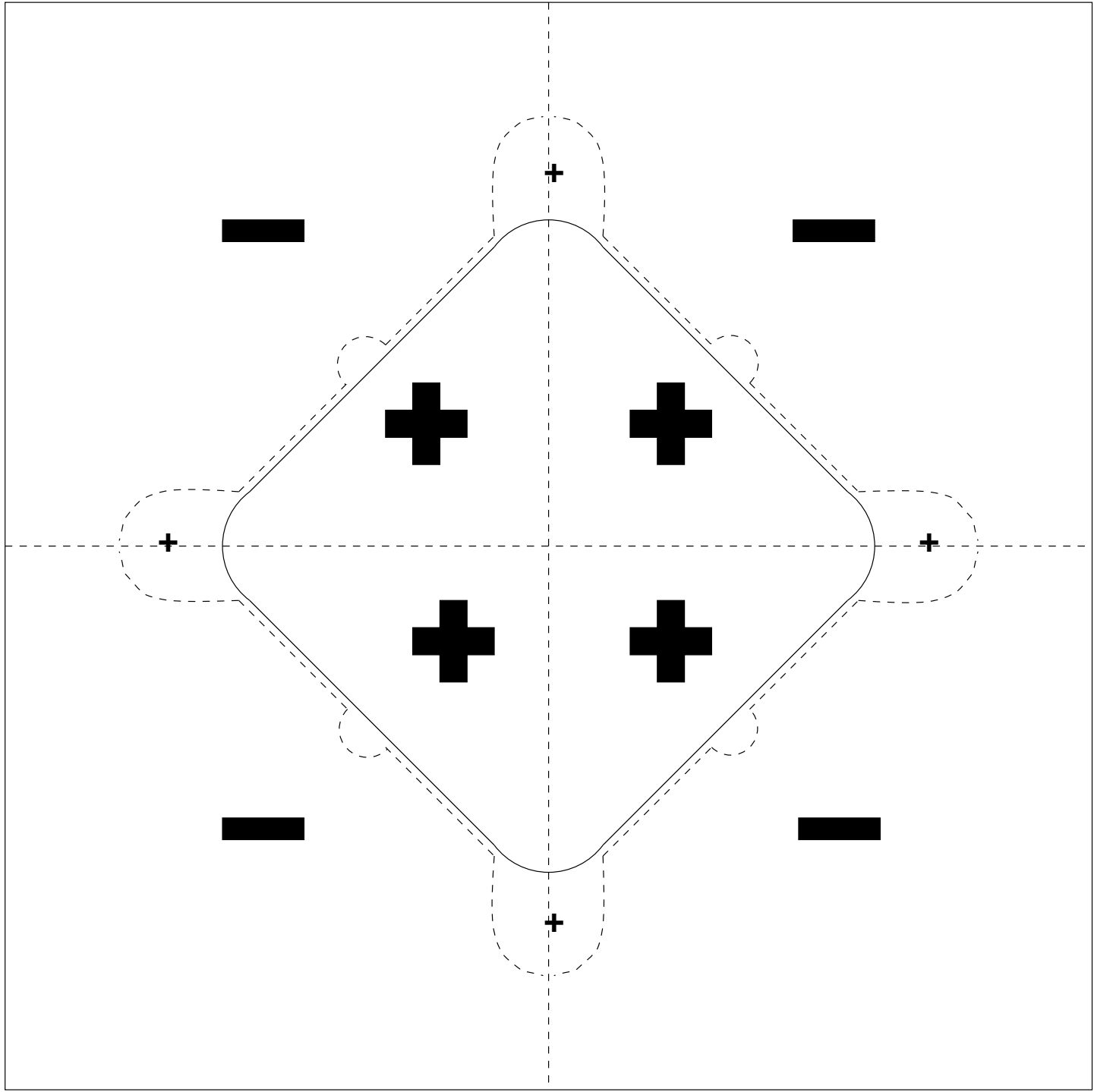
\end{document}